\documentclass[reprint,showpacs,preprintnumbers,amsmath,amssymb,prc,floatfix]{revtex4-1}
\usepackage{float}
\usepackage{color}
\usepackage{graphicx}
\usepackage{dcolumn}
\usepackage{bm}
\usepackage{xcolor}
\usepackage{multirow}
\usepackage{threeparttable}
\usepackage{adjustbox}
\usepackage{longtable}
\usepackage{comment}
\usepackage{natbib}
\usepackage{array}
\usepackage[utf8x]{inputenc}
\usepackage[colorlinks,citecolor=blue,linkcolor=red,anchorcolor=blue,filecolor=blue,urlcolor=blue]{hyperref}
\begin{document}
\title{Structure and reaction study of Z=120 isotopes using non-relativistic and relativistic mean-field formalism}
\author{Jeet Amrit Pattnaik$^1$}
\email{jeetamritboudh@gmail.com}
\author{K. C. Naik$^1$}
\email{kishornaik.physics@gmail.com}
\author{R. N. Panda$^{1}$}
\email{rnpandaphy@gmail.com}
\author{M. Bhuyan$^{2}$}
\email{bunuphy@um.edu.my}
\author{S. K. Patra $^{4,5}$}
\email{patra@iopb.res.in}
\affiliation{$^1$Department of Physics, Siksha 'O' Anusandhan, Deemed to be University, Bhubaneswar-751030, India}
\affiliation{$^2$Center for Theoretical and Computational Physics, Department of Physics, Faculty of Science, University of Malaya, Kuala Lumpur 50603, Malaysia.}
\affiliation{$^3$Institute of Physics, Sachivalya Marg, Bhubaneswar-751005, India}
\affiliation{$^4$Homi Bhabha National Institute, Training School Complex, Anushakti Nagar, Mumbai 400094, India}


\date{\today}

\begin{abstract}
An extensive study is carried out for the island of stability in the superheavy nuclei of Z=120 and N=164-220 within the effective field theory motivated relativistic mean-field (E-RMF) and the non-relativistic Skyrme-Hartree-Fock (SHF) approaches. The relativistic G3 and IOPB-I and non-relativistic SLy4 and SkMP parameter sets are used for the investigations. Surface properties such as symmetry energy, neutron pressure and the curvature coefficient of symmetry energy are discussed within the coherent density fluctuation model (CDFM) using the Skyrme and the Br\"uckner energy density functionals. The volume and surface contributions of symmetry energy are evaluated using Danielewicz's liquid drop approximation within the CDFM. The total nuclear reaction and elastic differential cross-sections are also obtained for both SHF and E-RMF within the Glauber model. The peaks in the symmetry energy at N = 182 for SHF and N=184 for E-RMF are seen, which are absent in the Br\"uckner functional. The shifting of peak in the symmetry energy with Br\"uckner functional can be correlated  to the Coester-band problem. The enhanced total reaction cross-section for relativistic density of $^{304}120$ suggests the extra stability of this nucleus. This further confirms the shell/sub-shell closure of N = 184 in E-RMF force. The differential cross-section shows its force independent nature and significant increase with the scattering angle.
\end{abstract}
\maketitle

\label{intro} 
\noindent
\section{Introduction} \noindent 
In the last few decades, more than 8000 nuclei have been predicted by the different mass models, which are suppose to be produced in the laboratories \cite{Naz98, Tho04}, where most of them are situated at the neutron-rich side of the nuclear landscape. Due to their unusual neutron to proton ratio, it is very difficult to produce in the laboratory \cite{Og06, Og12}. In the present era, the arrival of the notable radioactive ion beam techniques helps the nuclear researchers to explore more in detail about the drip-line as well as the super-heavy region.  A lot of physics and unanswered questions are there within the island of the superheavy elements (SHE) that physicists are desired to know nowadays. Out of which the magic number concept, as well as the double shell close of proton and neutron number, has an important position in the field of nuclear structure and reaction physics. The fighting between the strong attractive shell corrections against the repulsive Coulomb interaction decides the stability of SHE.\\\\
Using the Skyrme interaction, Kruppa {\it et. al.} \cite{Kar12} predicted relatively large  shell gaps for Z=124 and 126 with the neutron number  N=184 and predicted as possible close shell nuclei. However,  due to the difference in shell structure in E-RMF model, a contrasting result is studied at Z=120 and N=172.  Bhuyan {\it et. al.} \cite{Bhu12} by the help of non-relativistic Skyrme-Hartree-Fock (SHF) and the relativistic mean-field formalism suggests that Z=120 with N=172, 182/184, 208, 258 are the neutron magic numbers in the island of SHE after N=126. Also, the nuclei having  Z = 120 and N = 172, 258 are found to be spherical doubly magic super heavy nuclei as suggested by Sil {\it et. al.} \cite{Sil04}. Similarly, Biswal {\it et. al.} \cite{Bis14} suggested that nuclei having  Z=114, 120, 126 and N=182 are the magic nuclei with the help of simple effective interaction. T. Sahoo {\it et. al.} \cite{Sah18} used the density-dependent RMF method and observed that the arrangement with Z=114, 120, 126 and N=172, 184, 198 are magic combinations. The recent works bring lots of curiosity after the formation of three new elements with Z=110, 111, 112 \cite{Hof95p, Hof95, Og99} which address the zone of spherical superheavy magic nuclei at Z=114. However, no such magic properties are observed in this nucleus. This finding encourages further investigations to find the next magic one after $^{208}Pb$ and thought about the nucleus with Z=126. Because of Coulomb repulsion as well as the shell repulsion, this may not happen so. From the above analysis, one can notice the most possible candidate for a proton magic at Z = 120. In the case of neutron magic, we do not have any consistency in the theoretical predictions. In this direction, we have considered Z=120 with different combinations of neutron N = 164 – 220 within both the nuclear structure and reaction domain. Here, the motivation for the present work is to co-relate the nuclear structure properties to the nuclear reaction phenomena at shell/sub-shell closure with the help of various iso-spin dependent quantities, the so-called surface properties like symmetry energy, neutron pressure and symmetry energy curvature co-efficient etc. \\ \\
The neutron-proton (N-Z) asymmetry is controlled by the symmetry energy, which is referred  to be an iso-spin dependent quantity \cite{Bhu18}. It is consequential to test the available symmetry energy in order to analyse the structural properties of nuclei in this case because we are considering the drip line nuclei for this investigation. The coherent density fluctuation model (CDFM) which reconstructs  the infinite nuclear matter quantities at local density is used to compute the symmetry energy of a finite nucleus.
\\ \\
The search of magic numbers in the superheavy region can also be possible from the study of nuclear reaction phenomena and is equally important like the nuclear structural properties as mentioned above. The theoretical research works for the same open a path to be followed by the experimental physicists to answer some of the undefined problems in the unknown territory of the island of SHE. In this paper, superheavy nuclei with Z=120 is discussed with the basis of both nuclear structure and reaction dynamics which may guide for the new predictions of SHE with Z=120 as a magic nucleus experimentally in the superheavy region. 
\\
\\
The paper is structured as follows: in section \ref{theory1}, we briefly outline the formalism for the calculation of the ground state properties as well as the surface properties of a finite nucleus. The densities used are obtained from non-relativistic Skyrme-Hartree-Fock formalism with SLy4 and SkMP forces and Effective-relativistic mean-field theory with recent and most successful G3 and IOPB-I parameters, which are also discussed precisely in this work. The surface properties such as symmetry energy, neutron pressure and curvature coefficient of symmetry energy are also discussed within the coherent density fluctuation model (CDFM) along with the Br\"uckner energy density functional approach. The Dainelewicz's liquid drop approximation is also utilised for the account of the surface and volume contributions part of symmetry energy, which is explained in this section too. Here, we have implemented the newly derived energy density functionals for the used skyrme forces SLy4 and SkMP, for the first time. A fine fitting procedure is used in order to solve the issues with the saturation property and Coester band problem. Again the credibility of the newly derived Skyrme energy density functionals is confirmed by analysing the Dainelewicz's liquid drop approximation approach. In section \ref{results}, all the results and discussions are discussed. In this section, we have discussed the ground state properties of the considered nuclei for the chosen forces, including binding energy, charge distribution radius, neutron skin thickness, density profiles with corresponding weight functions, two neutron separation energy, neutron/proton pairing gap, and iso-spin dependent surface properties. We have also attempted to connect these structural properties with the reaction dynamics using measures like the total nuclear reaction and differential elastic scattering cross-sections.  The summary and concluding remarks are analysed in section \ref{conclusion}.

\section{Theoretical Framework}
\label{theory1}
\subsection{The Skyrme Hartree-Fock Method} 
The Skyrme effective interaction, used in the nuclear mean-filed models is one of the oldest and well documented formalism. For completeness, we only outline the essential expressions in this section. The Skyrme energy density functional is written as $\mathcal{H}$ \cite{Cha97,Cha98,Sto07},
\begin{eqnarray}
\mathcal{H}=\mathcal{K}+\mathcal{H}_0+\mathcal{H}_3+\mathcal{H}_{eff}+......
\end{eqnarray}
Here, $\mathcal{K}=\frac{\hbar^2}{2M}\tau$ is the kinetic energy with $M$ is the mass of the nucleon and $\tau=\tau_{p}+\tau_{n}$ is the kinetic energy density. The other terms $\mathcal{H}_0$, $\mathcal{H}_3$ and $\mathcal{H}_{eff}$ are the zero range, the density dependent and the effective-mass
dependent terms, respectively. There are nine parameters, $t_i$,
$x_i$ (i = 0, 1, 2, 3), and $\eta$ are used to calculate the relevant properties of finite nuclei and nuclear matter, which are given as:
\begin{eqnarray}
\mathcal{H}_0&=&\frac{1}{4}t_0\left[\left(2+x_0\right)\rho^2
-\left(2x_0+1\right)\left(\rho_p^2+\rho_n^2\right)\right]\\
\mathcal{H}_3&=&\frac{1}{24}t_3\rho^\eta\left[\left(2+x_3\right)\rho^2
-\left(2x_3+1\right)\left(\rho_p^2+\rho_n^2\right)\right]\\
\mathcal{H}_{eff}&=&\frac{1}{8}\left[t_1\left(2+x_1\right)
+t_2\left(2+x_2\right)\right]\tau\rho \nonumber\\
&+&\frac{1}{8}
\left[t_2\left(2x_2+1\right)
-t_1\left(2x_1+1\right)\right]\left(\tau_p\rho_n+\tau_n\rho_p\right).
\end{eqnarray}
The terms $\mathcal{H}_{S\rho}$ and $\mathcal{H}_{S\vec{J}}$ represent the surface contributions of a nucleus with the additional parameters $b_4$ and $b_4^\prime$ are written as:
\begin{eqnarray}
\mathcal{H}_{S\rho}&=&\frac{1}{16}\left[3t_1\left(1+\frac{1}{2}x_1\right)
-t_2\left(1+\frac{1}{2}x_2\right)\right]\left(\vec{\nabla}\rho\right)^2 \nonumber\\
&&-\frac{1}{16}\left[3t_1\left(x_1+\frac{1}{2}\right)
+t_2\left(x_2+\frac{1}{2}\right)\right]\nonumber\\
&&\times\left[\left(\vec{\nabla}\rho_n\right)^2+
\left(\vec{\nabla}\rho_p\right)^2\right],
\end{eqnarray}
and
\begin{equation}
\mathcal{H}_{S\vec{J}}=-\frac{1}{2}\left[b_4\rho\vec{\nabla}\cdot\vec{J}
+b_4^\prime\left(\rho_n\vec{\nabla}\cdot\vec{J_n}+
\rho_p\vec{\nabla}\cdot\vec{J_p}\right)\right].
\end{equation}
Here, $\rho_n$, $\rho_p$ and $\rho=\rho_n+\rho_p$ are the neutron, proton and total densities, respectively of the nucleus. The kinetic energy density and the spin-orbit density are written as $\tau=\tau_n+\tau_p$ and $\vec{J}=\vec{J_n}+\vec{J_p}$, respectively. The $\vec{J_q}=0$ ($q=n$ or $p$) for spin-saturated nucleus. The integral of $\mathcal{H}$ gives the total binding energy of the nucleus. The Skyrme parameter sets SLy4 and SkMP are used in our calculations with $b_4\neq b_4^\prime$~ \cite{Sto07,Sto03,Rei95}. These sets are constructed to reproduce a proper spin-orbit potential for finite nucleus and also the observed isotopic shifts in the Pb isotopes \cite{Pan11,Sha13,Ahm12,Pat07,Gup07,jeetchin}. \\ 
\subsection{Relativistic mean field formalism}
The effective field theory motivated relativistic mean-field (E-RMF) Lagrangian is outlined briefly  \cite{Fur96-97}. The renormalization and divergence of the Lagrangian that was the fundamental problems earlier in the formalism overcame, because of this effective Lagrangian. The straightforward coupling constants and few meson masses fitting made this formalism a very useful one. For real practice, a truncation method is needed as this E-RMF Lagrangian possesses infinite terms having self- and cross-couplings interactions \cite{Est01, Mul96, Ser97}. The nucleon-meson E-RMF Lagrangian up to a maximum of fourth-order terms for relevant mesons are fixed applying the naturalness and naive dimensional analysis. We get the energy density functional  ${\cal{E}}(k)_{nucl}$ as a function of scalar and vector density $\rho_s$ and $\rho_v$ respectively from the E-RMF Lagrangian,  which is written as \cite{Sin14, Fur87, Kum18}:
\begin{eqnarray}
{\cal{E}}(k)&=&\frac{2}{(2\pi)^{3}}\int d^{3}k E_{i}^\ast (k)+
\frac{ m_{s}^2\Phi^{2}}{g_{s}^2}\Bigg(\frac{1}{2}+\frac{\kappa_{3}}{3!}
\frac{\Phi }{M} 
\nonumber\\
&&
+ \frac{\kappa_4}{4!}\frac{\Phi^2}{M^2}\Bigg)
+\rho_b W-\frac{1}{4!}\frac{\zeta_{0}W^{4}}
{g_{\omega}^2}
-\frac{1}{2}m_{\omega}^2\frac{W^{2}}{g_{\omega}^2}
\nonumber\\
&&
\Bigg(1+\eta_{1}\frac{\Phi}{M}+\frac{\eta_{2}}{2}\frac{\Phi ^2}{M^2}\Bigg)
+\frac{1}{2}\rho_{3}R
-\frac{1}{2}\Bigg(1+\frac{\eta_{\rho}\Phi}{M}\Bigg)
\nonumber\\
&&
\frac{m_{\rho}^2}{g_{\rho}^2}R^{2}
-\Lambda_{\omega}  (R^{2}\times W^{2})
+\frac{1}{2}\frac{m_{\delta}^2}{g_{\delta}^{2}}D^{2}.
\label{enm}
\end{eqnarray}
In the above Eq. \ref{enm}, the $\Phi$, $W$, $R$ and $D$ are the desired fields for $\sigma$, $\omega$, $\rho$ and $\delta$ mesons prescribed as $\Phi = g_s\sigma$, $W = g_\omega \omega^{0}$, $R = g_\rho \vec{\rho}^{0}$ and $D=g_{\delta}\delta^{0}$, respectively. Similarly, the masses of nucleon, $\sigma$, $\omega$, $\rho$ and $\delta$ mesons are defined as $M$, $m_{\sigma}$, $m_{\omega}$, $m_{\rho}$ and $m_{\delta}$ respectively \cite{Ank21}. 
\subsection{Fitting procedure}
The fitted binding energy function for E-RMF and SHF is expressed as \cite{Ank21}:
\begin{eqnarray}
{\cal E}(x) & = & C_k \rho_0^{2/3}(x) + \sum_{i=3}^{14} (b_i + a_i \alpha^2) \rho_0^{i/3}(x).
\label{efitting}
\end{eqnarray}
The first term expresses the kinetic energy having coefficient $C_k$ is given as $C_k = 37.53 [(1+\alpha)^{5/3} + (1-\alpha)^{5/3}]$, following the Thomas-Fermi approach. Here, the $\alpha$ can be written as n-p assymetry parameter ($\alpha = \frac{\rho_n - \rho_p}{\rho_n + \rho_p}$) for E-RMF and for SHF, $\alpha=1-2*y_p$, where $y_p$ is the proton asymmetry. Kumar {\it et. al.} \cite{Ank21} have considered 12 terms as the best fit and the coefficients has been extrapolated from the polynomial fitting. A detailed procedure can be found in \cite{Ank21}. \\
The NM quantities are as follows:
\begin{eqnarray}
S^{NM} &=& 41.7\,\rho_0^{2/3}(x) + \sum_{i=3}^{14} a_i\, \rho_0^{i/3}(x), \label{eqB}\\
P^{NM} &=& 27.8\,\rho_0^{5/3}(x) + \sum_{i=3}^{14} (i/3)\, a_i\, \rho_0^{(i+3)/3}(x), \label{eqC}\\
K_{sym}^{NM} &=& -83.4\,\rho_0^{2/3}(x) + \sum_{i=4}^{14} i\, (i-3)\, a_i\, \rho_0^{i/3}(x). \label{eqD}
\end{eqnarray}
\subsection{Br\"{u}ckner's prescription and Symmetry Energy}\label{fitting}
The nuclear matter symmetry energy $S^{NM}$, $P^{NM}$ and $K_{sym}^{NM}$ are derived from the well known Br\"{u}ckner energy density functional can be established as \cite{Bur68,Bur69,Kum18,Che14,Pat21a,Pat21b}:
\begin{eqnarray}
S^{NM} &=& 41.7\rho_0(x)^{2/3}+b_4\rho_0(x)+b_5\rho_0(x)^{4/3}
\nonumber\\
&& + b_6\rho_0(x)^{5/3},\label{snm} \\ 
P^{NM} &=& 27.8\rho_0(x)^{5/3}+b_4\rho_0(x)^2+\frac4 3 b_5\rho_0(x)^{7/3} \nonumber\\
&& +\frac5 3 b_6\rho_0(x)^{8/3}, \label{pnm} \\
K_{sym}^{NM} &=& -83.4\rho_0(x)^{2/3}+4b_5\rho_0(x)^{4/3} \nonumber\\
&& +10b_6\rho_0(x)^{5/3}.  
\label{ksymn}
\end{eqnarray}
The calculated nuclear densities of Z=120 nuclei from RMF feeded as input in Coherent Density Fluctuations Model (CDFM).  \\
\subsection{Coherent Density Fluctuations Model} \label{CDFM}
The Coherent Density Fluctuations Model (CDFM) \cite{Ant80} is a way to find symmetry energy and its derivatives, by obtaining the NM expressions within a local density approximation (LDA), combined with the weight function derived from finite nuclear densities \cite{Ant80, Ant4, Ant2, Ant3}. Details can be found in references \cite{Pat21a,Pat21b,Bhu18,Gad11}. \\
\\
The weight function $|F(x)|^2$ for a density $\rho$ (r) of a nucleus having mass number A, derived with relativistic mean field model can be defined as:
\begin{equation}
|F(x)|^2 = - \left (\frac{1}{\rho_0 (x)} \frac{d\rho (r)}{dr}\right)_{r=x},
\label{wfn}
\end{equation}
with $\int_0^{\infty} dx \vert F(x) \vert^2 =1$ \cite{Ant4, Ant2, Gad11, Bhu18, Gad12}. The finite nuclear symmetry energy $S^{A}$ is calculated by weighting the corresponding quantity for infinite NM within the CDFM, as given below \cite{Ant4, Gad11, Gad12, Ant17,Fuc95}
\begin{eqnarray}
S^{A}= \int_0^{\infty} dx\, \vert F(x) \vert^2\, S^{NM} (\rho (x)) ,
\label{s0}
\end{eqnarray}
\begin{eqnarray}
P^{A} =  \int_0^{\infty} dx\, \vert F(x) \vert^2\, P^{NM} (\rho (x)),
\label{p0}
\end{eqnarray}
\begin{eqnarray}
K_{sym}^{A} =  \int_0^{\infty} dx\, \vert F(x) \vert^2 \ K_{sym}^{NM} (\rho (x)).
\label{k0}
\end{eqnarray}
The $S^{A}$, $P^{A}$, $K_{sym}^{A}$ in  Eq. (\ref{s0}-\ref{k0}) are the surface weighted average of the corresponding nuclear matter quantity at local density for finite nuclei.\\
\subsection{Volume and surface symmetry energy in Danielewicz's liquid drop prescription}
The volume $S_V$ and surface $S_S$ contributions of symmetry energy can be obtained as \cite{Dan03,Dan04, Dan06}:
\begin{eqnarray}
S_{V}=  S^A \left (1+\frac{1}{\kappa A^{1/3}} \right)
\end{eqnarray}
and
\begin{eqnarray}
S_{S}= \frac {S^A}{\kappa} \left (1+\frac{1}{\kappa A^{1/3}} \right).
\end{eqnarray}
Details can be found in references \cite{Bhu18,Ant2,Ant80,Pat21a,Pat21b}. The ratio $\kappa \equiv \frac{S_V^{A}}{S_S^{A}}$ with $\gamma$ = 0.3 \cite{Ant18} is described below;
\begin{eqnarray}
\kappa = \frac{3}{R\rho_{0}}\int_0^{\infty} dx \vert F(x) \vert^2 x \rho_{0}(x)
\left (\left (\frac{\rho_{0}}{\rho(x)}\right )^{\gamma}-1\right).
\label{k0t}
\end{eqnarray}. 
\subsection{Glauber model}
The Glauber model is successfully used from near the Coulomb barrier to higher energy region for the study of heavy-ion collision to explain many interaction phenomena. Within a fixed density profile the nucleus-nucleus interaction is provided by this formalism. This method is a semi-classical concept where the impact parameter explains the nuclear collision when the nuclei are in collision direction along a straight line. The Glauber formalism provides satisfactory results for both the differential elastic cross-sections and the nucleus–nucleus reaction over a wide range of energies \cite{Gla59}. \\
\subsubsection{Total nuclear reaction cross section}
The Glauber model \cite{Gla59} along the eikonal approximation calculate the total nuclear cross-section $\sigma_r$. Here, the nucleon-nucleon ($NN$) collisions are considered as independent and individual event \cite{Abu03}. The original form of the total reaction cross-section $\sigma_r$ of the Glauber model at high energies is given as \cite{Gla59, Kar75}:
\begin{equation}
\sigma_r=2\pi\int_0^\infty \textbf{b}[1-T(\textbf{b})]d\textbf{b},
\end{equation}
here T(\textbf{b}) is the Transparency function and \textbf{b} is the impact parameter. Although the original Glauber model was designed for high energy approximation but it works quite well for both the nucleus-nucleus reaction and the differential elastic cross-sections over a broad energy range of incident energy \cite{cha83,buenerd1984}. \\
\\
The axially deformed densities obtained from the E-RMF/SHF models can not be used directly in cross-section study, so we have used the spherical equivalent of the deformed E-RMF or spherical SHF densities, which are fitted to a sum of two Gaussian functions to obtain suitable co-efficient $c_i$ and ranges $a_i$, in our calculations \cite{Abu03}:
\begin{equation}
\rho (r)=\sum\limits_{i=1}^{2}c_{i}exp[-a_{i}r^{2}].
\label{rhor}
\end{equation}
Here $c_i$ and $a_i$ are the Gaussian co-efficients fitted with the $z$-integrated density, which is defined as:
\begin{equation}
\overline{\rho}(w) =\int_{-\infty}
^\infty\rho(\sqrt{w^{2}+z^{2}}) d\textbf{z}, 
\label{equrho}
\end{equation}
with $w^2=x^2+y^2$. 
The above co-efficients are used to calculate the total reaction cross-section for both the stable and unstable nuclei considered in the present study. In refs. \cite{Pan09,Abu03,Shu03,bhagwat}, it is shown that the Glauber model can be used for relatively low energy even at 25, 30 and 85 MeV/nucleons.
\subsubsection{Angular elastic differential cross section}
The elastic scattering amplitude for a nucleus-nucleus interaction is written as,
\begin{equation}
F(\textbf{q})=\frac{\iota K}{2\pi} \int d\textbf{b} e^{\iota \textbf{q}.\textbf{b}}(1-e^{\iota \chi_{PT}(\textbf{b})}).
\end{equation}
This model is modified to take care of finite range effects in the profile function and Coulomb modified trajectories at low energy. The elastic 
scattering amplitude along with the Coulomb interaction is expressed as
\begin{equation}
F(\textbf{q})=e^{\iota \chi_{s}}\{F_{coul}(\textbf{q})+\frac{\iota K}{2\pi}
\int d\textbf{b} e^{\iota \textbf{q}.\textbf{b}+2\iota \eta \ln(Kb)}(1-e^{\iota \chi_{PT}(\textbf{b})})\},
\end{equation}
and the Coulomb elastic scattering amplitude
\begin{equation}
F_{coul}(\textbf{q})=\frac{-2 \eta K}{q^2}exp\{-2 \iota \eta \ln(\frac{q}{2K})
+2\iota arg \Gamma(1+\iota \eta)\}.
\end{equation}
Here $K$ is the momentum of projectile and $q$ is the momentum transferred from the projectile to the target and $\eta=Z_P Z_T e^2/\hbar v$ is the Sommerfeld parameter, $v$ is the incident velocity of the projectile, and $\chi_s=-2\eta \ln(2 K a)$ with $a$ being the screening radius.The elastic differential cross section is given by
\begin{equation}
\frac{d\sigma}{d\Omega}=|F(\textbf{q})|^2.\\
\end{equation}
The ratio of angular elastic to the Rutherford elastic differential cross section is written as:
\begin{eqnarray}
\frac{d\sigma}{d\sigma_r}=\frac{\frac{d\sigma}{d\Omega}}{\frac{d\sigma_r}{d\Omega}}=\frac{|F(\textbf{q})|^2}{|F_{coul}(\textbf{q})|^2}.
\end{eqnarray}
\begin{table*}
\centering
\caption{The nuclear matter properties at saturation, i.e.,  saturation density ($\rho_{0}$), binding energy per nucleon ($E_0$), effective mass ($M^*$/M), symmetry energy ($J$), slope parameter ($L$), second ($K_{sym}$) and third ($Q_{sym}$) order derivative of symmetry energy and incompressibility ($K_{\infty}$) for SLy4, SkMP, IOPB-I and G3 parameter sets. The empirical/experimental values are also given for comparison. The $\rho_0$ is in fm$^{-3}$, $\frac{M*}{M}$ is dimensionless and  all other quantities are in MeV.}

\begin{tabular*}{\linewidth}{c @{\extracolsep{\fill}}cccccc}
\hline
\hline
Parameters& G3 & IOPB-I & SLy4 & SkMP & Empirical Value &  \\ \hline
$\rho_{0}$ & 0.148 & 0.149 & 0.160 & 0.157 & (0.148)-(0.185) \cite{Bethe}   &  \\
$E_0$  & -16.02 & -16.10 & -15.97 & -15.56 & (-15.0)-(-17.0) \cite{Bethe} & \\
$\frac{M^*}{M}$   & 0.699 & 0.593 & 0.69  & 0.65 & (0.55)-(0.60) \cite{Mark}&\\
$J$  & 31.84 & 33.30 & 32.00 & 29.89 & (30.00)-(33.70) \cite{Dani}& \\
$L$  & 49.31 & 63.58 & 45.94 & 70.31 & (35.0)-(70.0) \cite{Dani}&  \\
$K_{sym}$ & -106.07 & -37.09 & -119.73 & -49.82 & (-174.0)-(31.0) \cite{Zim}            &  \\
$Q _{sym}$ & 915.47 & 862.70 & 521.53  & 159.44 & (-494)-(-10) \cite{Cai}            &  \\
$K_{\infty}$ & 243.96 & 222.65 & 229.91 & 230.87 &  (220)-(260) \cite{Garg} &  \\
\hline     
\hline     
\end{tabular*}
\label{bulkproperties}
\end{table*}
\subsection{Choice of parameter sets}
To study the different nuclear properties, various parameter sets ($\sim265$) are designed \cite{Dut14, Kum17}. Each coupling term utilized in the effective field theory motivated relativistic mean-field (E-RMF) formalism has it's physical significance to elaborate certain concepts. Mesons have no self-interaction in the case of linear $\sigma-\omega$ model (Walecka model) \cite{Wal74}, which is the cause that the nuclear matter incompressibility parameter $K_{\infty}$ is found to be very high (550 MeV) as compared to the experimental results $\sim240 ± 20$ MeV \cite{Garg}. The nuclear matter properties at saturation are given in Table \ref{bulkproperties} for all the four considered parameter sets. The empirical/experimental data are also depicted for comparison. To decrease the value of $K_{\infty}$ to an acceptable value, Boguta and Bodmer added the self-coupling terms in $\sigma$ meson potential \cite{Bog77, Fu57, Pie01}. After adding such a self-coupling term, the value of $K_{\infty}$  reaches to $\sim 210-280$ MeV at saturation for infinite nuclear matter, depending on the force parameter \cite{rein86,ring87,lala97}. In addition to this, the repulsive term of the NN potential is also generated at the long-range region along with the excellent prediction of finite nuclei properties with the experimental data \cite{Gam90,rein86,ring87,lala97,Kum17,Kum18}. However, the problem arises again when it is unable to reproduce the nuclear matter equation of state (EoS) \cite{danisc,aru04}. The EoS can be made soften as per the requirements with the help of the non-linear term of the isoscalar vector meson \cite{Sug94}.  In the beginning, effective field theory motivated relativistic mean-field (E-RMF) formalism given by Furnstahl {\it et. al.} \cite{Fur96-97}, the cross-coupling parts of $\omega$ and $\rho$ mesons have not been included considering it's negligible effects. However, Todd-Rutel and Piekarewicz \cite{Tod05, Hor01} realized the effect of this term on the neutron distribution radius and  the neutron-skin thickness as well as the radius of the neutron star. The $\omega$ mesons have a self-coupling term with the coupling constant $\zeta_0$ which may tune accordingly to get the maximum neutron star mass and the experimental results for the case of sub-saturation density to a good agreement. The most successful forces namely G3 and IOPB-I with SLy4 and SkMP sets of the non-relativistic Skyrme-Hartree-Fock interaction have been taken for the calculations in the present work. 

\section{Results and Discussions}
\label{results}
We have analyzed a variety of structural properties including binding energy, neutron, proton, and total density distributions, charge distribution radius, neutron skin thickness, two neutron separation energy, neutron and proton pairing gap etc. for Z=120 isotopes with neutron numbers ranging from N=164 to 220. We have also examined the surface properties like symmetry energy, neutron pressure, symmetry energy curvature coefficients etc. A comparision is given regarding the study of the surface properties  with the two different energy density functionals i.e previously established conventional Br\"{u}ckner prescription and newly derived skyrme energy density functional.  We have also estimated the total nuclear reaction and the differential elastic scattering cross-sections by using spherical/spherical-equivalent densities. The axially deformed densities are converted to the spherical equivalent in two Gaussian form-fitting for all the above isotopes at various incident energies and angular distributions respectively. The detailed results of our calculations are presented in subsequent subsections and Figures \ref{fig1}-\ref{fig9}, in which we have used both the non-relativistic (SLy4, SkMP) and the relativistic (G3, IOPB-I) parameter sets. Furthermore, details can be found in their respective subsequent sections.
\subsection{Ground state properties}
To gain a thorough structural behaviour of the isotopic chain of considered proton shell closure Z=120, first we obtain the bulk properties such as binding energy, charge distribution radius, two neutron separation energy, neutron skin thickness, etc. Again, it is obvious that, with the exception of the drip-line region, bulk features are highly helpful in determining the shell closures. In the drip-line region, where there is significant neutron-proton asymmetry (N-Z) appears, the bulk properties can give inaccurate answers that can be corrected by looking at surface parameters like symmetry energy, neutron pressure, symmetry energy curvature coefficients, etc.
\\
In this subsection, we present the ground state properties of the isotopes with atomic number Z=120. The binding energies and charge radii with non-relativistic and relativistic formalism are calculated. These results are compared with each other and found that these are in general comparable.  For example, the binding energy and charge radius for $^{304}$120 with non-relativistic (SLy4 and SkMP) and relativistic (G3 and IOPB-I) frames are (2107.6, 6.295), (2059.08, 6.327) and (2131.02, 6.338), (2135.69, 6.329) in MeV \& fm,  respectively. The binding energy with finite range droplet model (FRDM) \cite{moller} is  2139.18 MeV. These results are quite close to each other. Similarly, the charge radius for this nucleus with SLy4, SkMP, G3 and IOPB-I are also consistent with each other. A detailed comparison with relativistic and non-relativistic results can be found in Ref. \cite{Bhu12}.\\
\begin{figure}[H]
\centering
\includegraphics[width=1.0 \columnwidth]{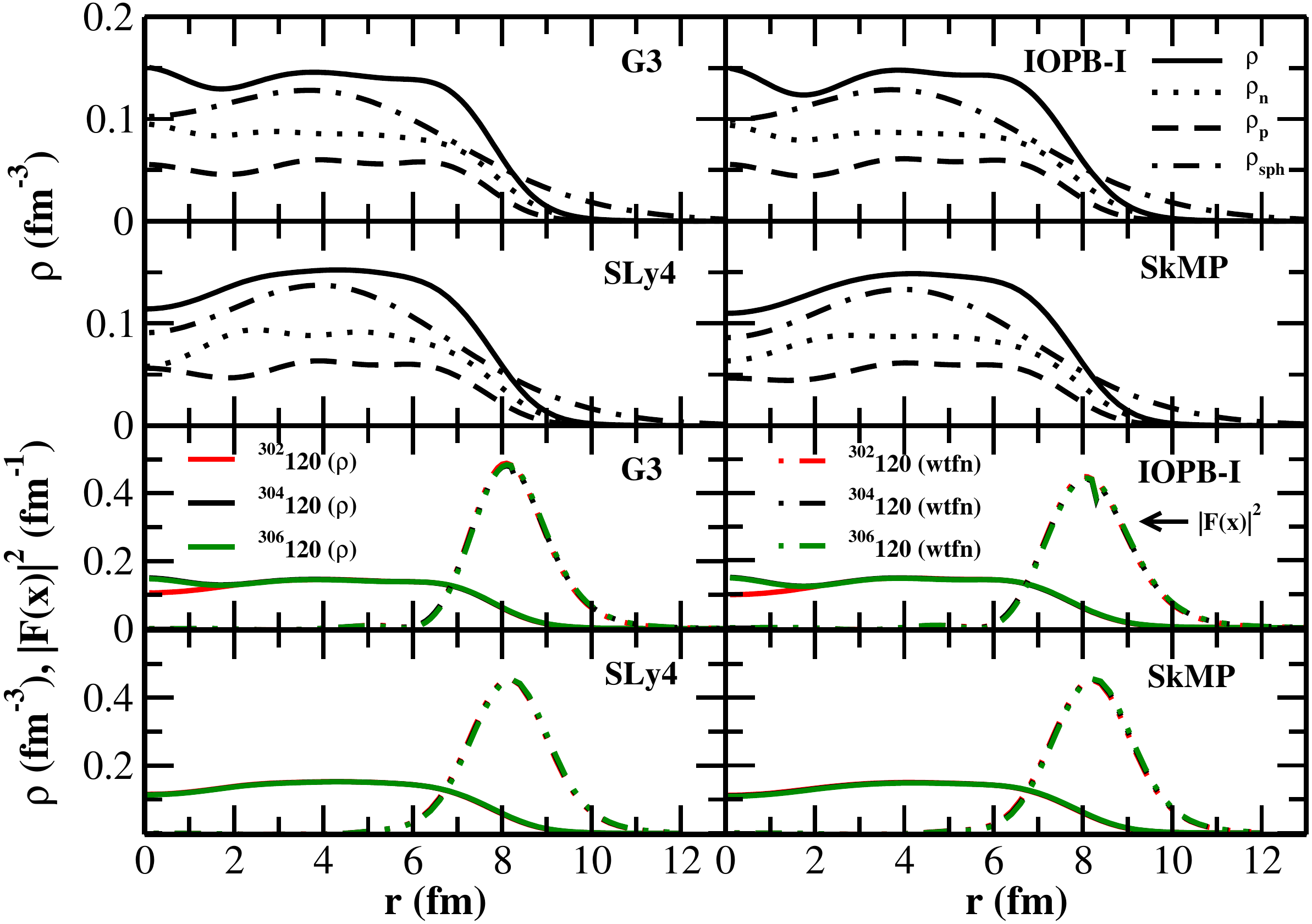}
\vspace*{8pt}
\caption{(Color online) In upper two panels: the neutron, proton, and total density distributions along with the respective spherical equivalent for Z=120 (N=184) with the relativistic forces (G3 and IOPB-I) and non-relativistic parameter sets (SLy4 and SkMP). The density is in $fm^{-3}$  and radius is in $fm$. In lower two panels: the total densities and weight functions ($|F(x)|^2$, dot-dashed line) for Z=120 (N=182, 184 and 186) with the relativistic and non-relativistic forces.} 
\label{fig1}
\end{figure}
The density calculated with any model is an acid test for its applicability to any nuclear system. The neutron is a neutral particle and its distribution inside the nucleus is very difficult to detect experimentally. On the other hand, due to the positive charge nature of the proton, the measurement of its distribution is almost accurate.  As a result, the experimental observation of charge radius $r_{ch}$ is also precisely measured. Thus, the theoretical density distributions of neutron give us enough inside about its distribution in a nuclear system and the comparison of charge density distribution with experiment gives us the applicability of the theoretical models. Since the observational data for density distribution of $^{304}$120 is not available, here we compare the results obtained from different parameter sets.\\

The ground state neutron, proton, and total density distributions for $^{304}$120 nucleus are shown in Figure \ref{fig1}  for relativistic G3, IOPB-I and non-relativistic SLy4, SkMP parameter sets. The central total ($\rho=\rho_n+\rho_p$) density in the non-relativistic case fails to attain its nuclear matter saturation (NM) density of $\rho_0=0.16$ fm$^{-3}$ suggesting a bubble-like structure. The NM saturation density $\rho_0$ obtains at $\sim 4$ fm for these two sets. In fact, a plateau-like structure of density distribution appears in the range $2-7$ fm as shown in the figure.  The $\rho$ distribution for the E-RMF case, however, shows up and down (wave-like structure) starting from the centre reflecting the shell structure of the nucleus. The central density is also almost equal to the NM saturation density ($\rho_0=$0.148 and 0.149) for this nucleus. We notice from Figure \ref{fig1}, that the neutron density for E-RMF (G3 and IOPB-I) are considerably different near the central region in comparison to non-relativistic cases (SLy4 and SkMP). Further, this difference reduces while moving towards the tail region.

Sometimes the spherical equivalent density $\rho_{sph}$ is essential for the evaluation of certain quantities. For example, in the calculation of total reaction cross-section $\sigma_r$ or the differential cross-section $\frac{d\sigma}{d\Omega}$ using Glauber model formalism \cite{Gla59, Abu03}. In this technique, the axially deformed density or even the spherically calculated density is converted to its spherical equivalent for getting an analytical form which is used in the numerical calculations.

Thus, we follow the procedure of Eqs. (\ref{rhor})-(\ref{equrho}) using a two Gaussian fitting and estimate the $c_1$, $c_2$ and $a_1$, $a_2$ to draw the spherical equivalent densities of the considered isotopes. As representative cases, we have plotted the $\rho_{sph}$ for the nuclei Z=120 with neutron numbers N=182, 184 and 186 which are represented in Figure \ref{fig1} as dashed-dot lines. The spherical equivalent densities are found to be lower in magnitude in the central region with an elongated tail. These densities will be used while calculating the $\sigma_r$ and the differential cross-section.

\subsection{The nuclear density and weight function}
\label{density}
In the lower two panels, the nuclear densities for some of the selected isotopes $^{302,304,306}$120 and their weight function ($|F(x)|^2$) with relativistic (G3 and IOPB-I) and non-relativistic (SLy4 and SkMP) formalism are plotted. The upper two panels of the plot show the densities and  their spherical equivalent $\rho_{sph}$. It is seen from the figure that the peak of the weight function is overlapped at the tail region of the density, i.e. the maximum portion of the weight function $|F(x)|^2$ is situated in the surface of the atomic nucleus. This is also noticed by Quddus {\it et. al.} in \cite{abduljpg} justifying the symmetry energy ($S^A$), neutron pressure ($P^A$) and the curvature coefficient of symmetry energy ($K_{sym}^A$) as the surface properties of the nucleus \cite{Pat21b}. These surface properties are calculated with the help of the CDFM approach at local density approximation, where the SHF and E-RMF densities are taken as input. The detailed procedure can be found in section \ref{CDFM}. Further, in Figure \ref{fig1}, we find the weight function is almost the same for the 3 considered nuclei. A minor/small change in the surface part of the density leads to a large variation in the expression of weight function and predict very different surface properties by folding with the nuclear matter quantities, defined in Eqs. (\ref{snm}) - (\ref{ksymn}).
\begin{figure}[H]
\centerline{\includegraphics[width=1.0 \columnwidth]{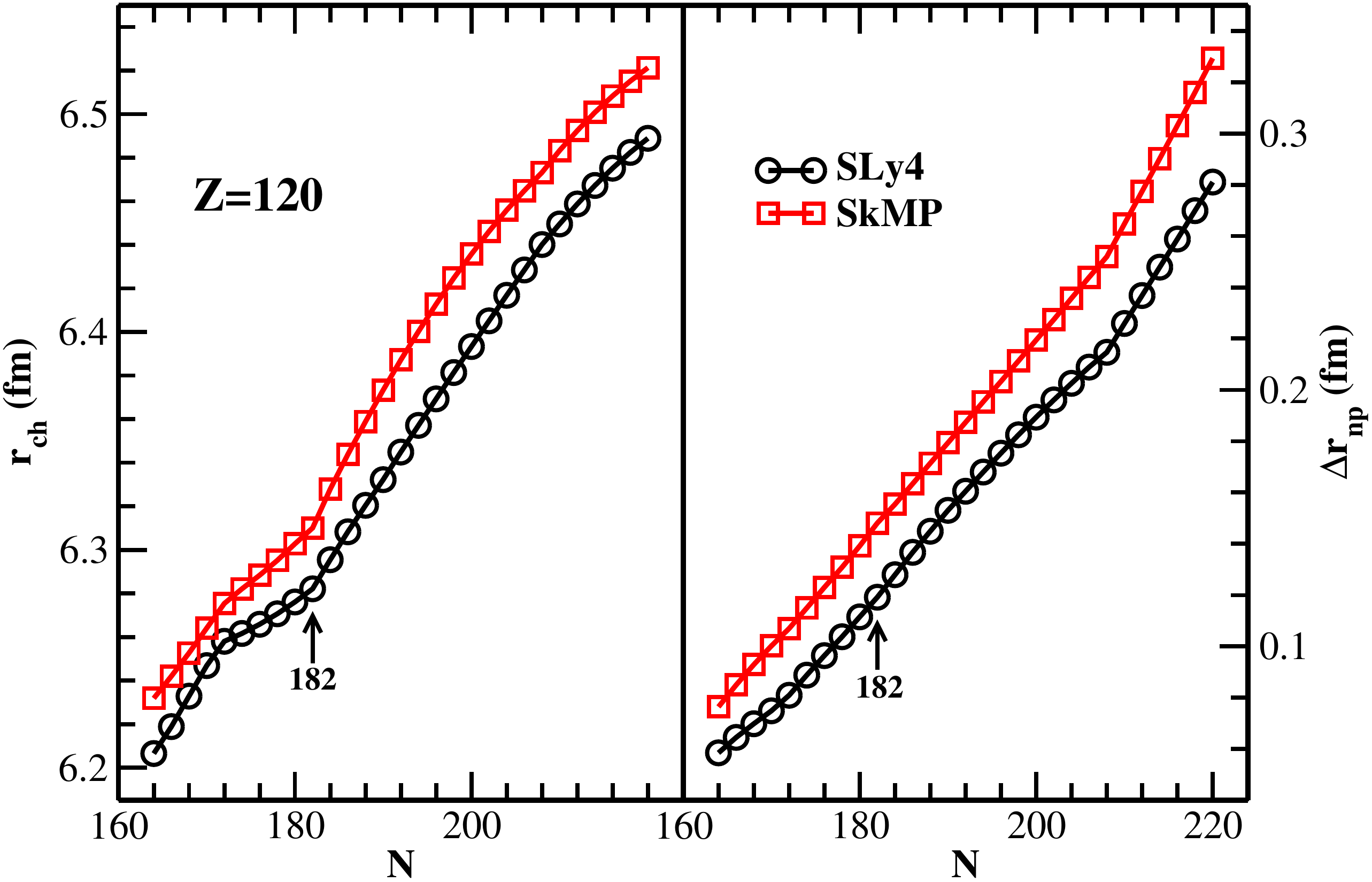}}
\vspace*{10pt}
\caption{(Color online) The charge distribution radius $r_{ch}$ and neutron skin thickness ($\Delta r_{np} = r_n-r_p$) are presented for Z=120 isotopic series with non-relativistic SLy4 and SkMP sets as a function of neutron number N. The possible magic number N=182 predicted with SHF is shown by an arrow.} \protect\label{fig2}
\end{figure}

\subsection{Nuclear charge radius and neutron-skin thickness}
The nuclear charge radius ($r_{ch}$) and the difference in root mean square radii between neutrons and protons distribution ($\Delta r_{np}=r_n-r_p$) are shown in Figure \ref{fig2}. The results, such as neutrons and protons distribution radii obtained from the self-consistent E-RMF and SHF calculations are in agreement with each other and a detailed discussion can be found in Ref. \cite{Bhu12}. Here, only the  $r_{ch}$ and  the neutron skin thickness $\Delta r_{np}$ are shown in Figure \ref{fig2} for SLy4 and SkMP parameter sets. The $r_{ch}$ and neutron-skin thickness increase monotonously with the mass number in the isotopic chain. The linear increase in neutron skin thickness suggests the surface/volume saturation that leads to an increase in the symmetry energy. Also, these results indicate the presence of more and more neutrons on the surface region of the nucleus. Further, a proper inspection shows that a depletion at N=182 is noticed for both the considered non-relativistic sets indicating the neutron magic number at N=182. This kink is found by Bhuyan {\it et. al.} for relativistic sets \cite{Bhu12}. Here, both the results obtained from SLy4 and SkMP forces follow the same trend. In the case of neutron skin thickness, the kinks are missing in non-relativistic cases however, these kinks with the relativistic G3 and IOPB-I sets are visible \cite{Pat21a}.
\subsection{Neutron separation energy and Pairing gap}
\begin{figure}[H]
\centerline{\includegraphics[width=1.0 \columnwidth]{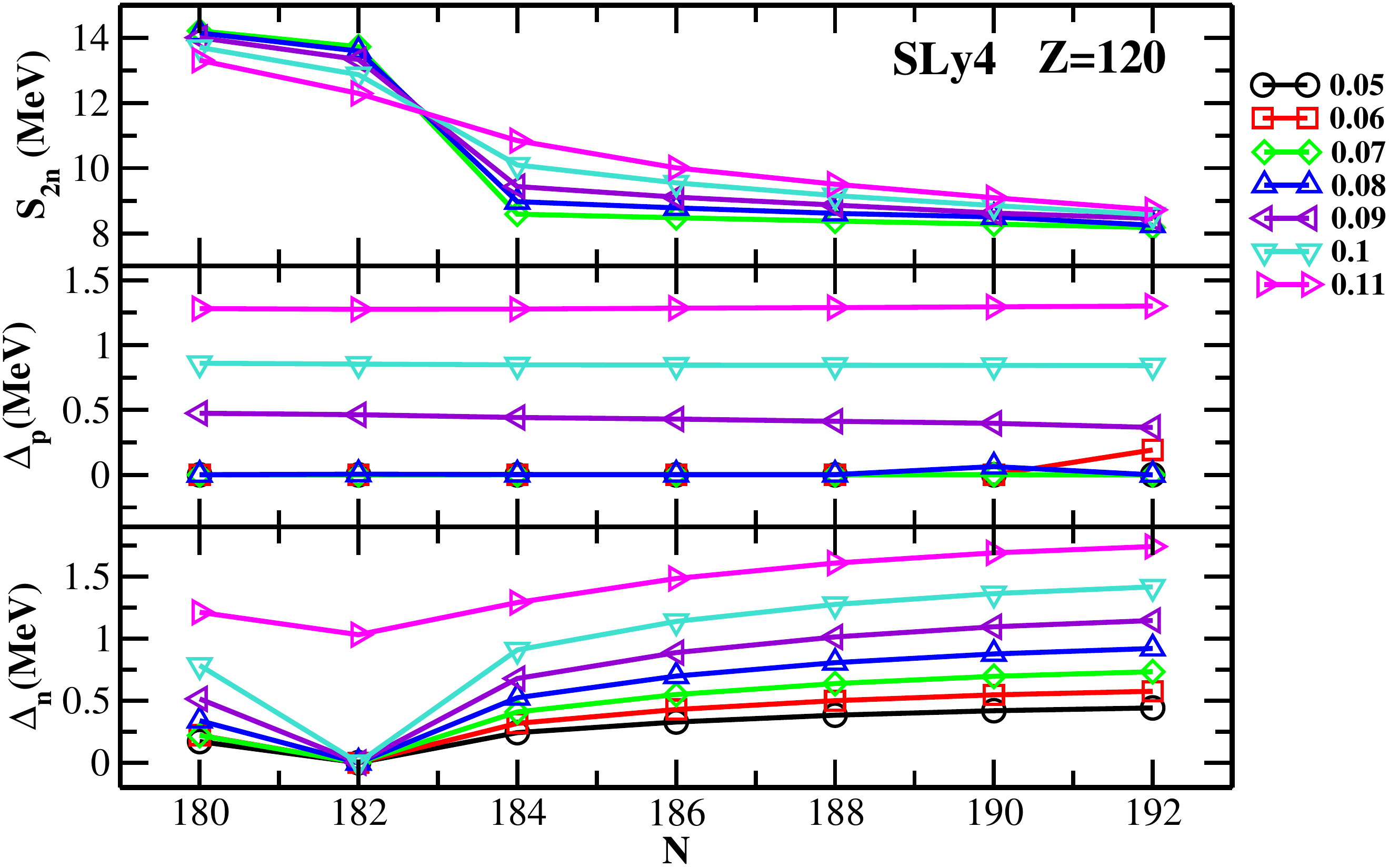}}
\vspace*{8pt}
\caption{(Color online) The two neutron separation energy ($S_{2n}$) (upper panel), proton pairing gap ($\Delta_p$) (middle panel), and neutron pairing gap ($\Delta_n$) (lower panel) are shown for neutron numbers N= 180-192 with the non-relativistic force parameters SLy4 and SkMP.} \protect\label{fig10}
\end{figure}
\begin{figure}[H]
\centerline{\includegraphics[width=1.0 \columnwidth]{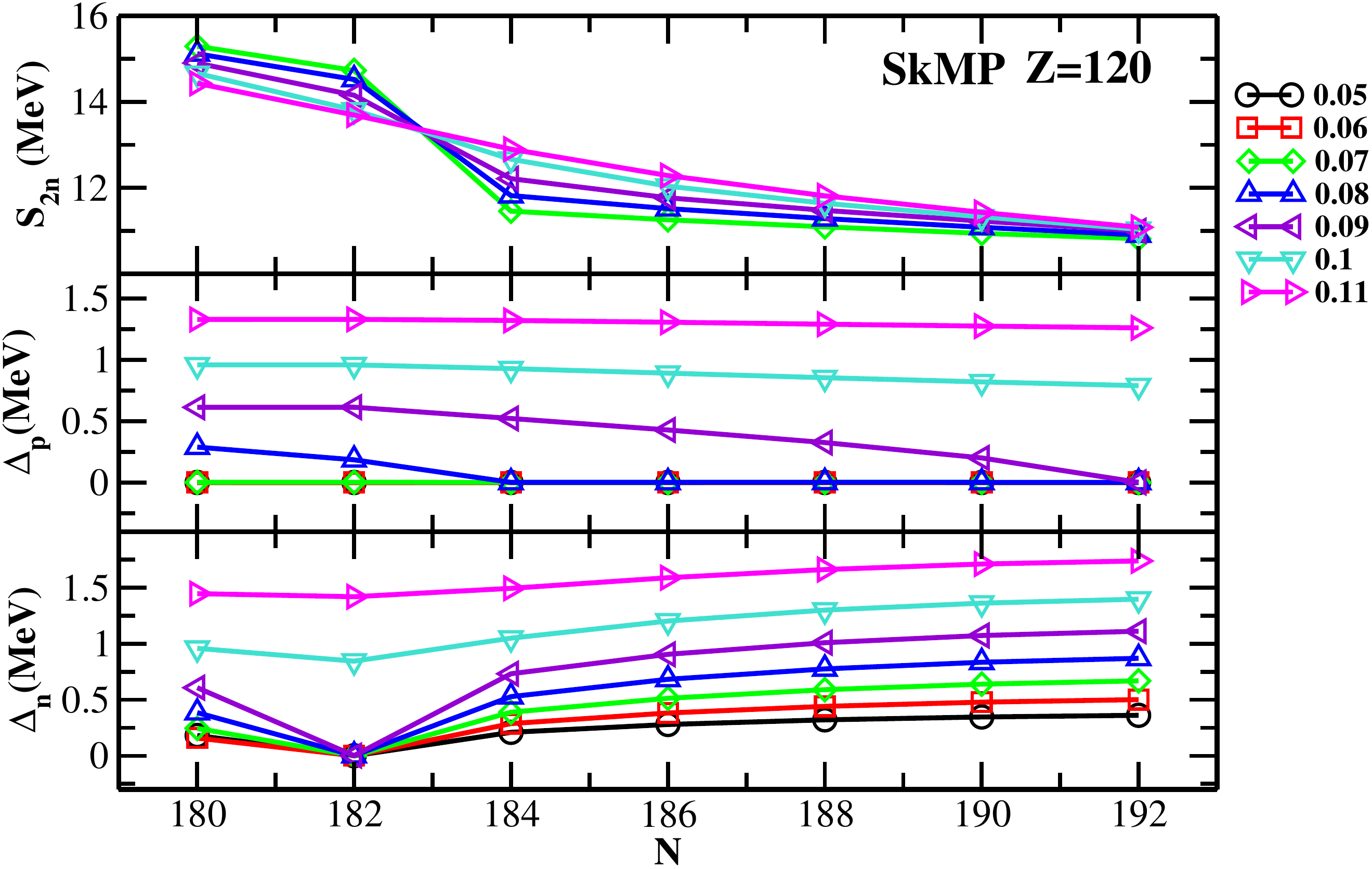}}
\vspace*{8pt}
\caption{(Color online) Same as Fig. \ref{fig10}, but for SkMP force.} \protect\label{fig11}
\end{figure}

For clarity point of view, in Figs. \ref{fig10} and \ref{fig11} we have shown the two neutron separation energy ($S_{2n}$) (upper panel), proton pairing gap ($\Delta_p$) (middle panel), and neutron pairing gap ($\Delta_n$) (lower panel) by analysing different inputs of pairing strength ranging from 0.05 to 0.11 for both the considered non-relativistic forces. In the relativistic case, the same can be found in our previous work \cite{Pat21a, Bhu12}. Here, for both SLy4 and SkMP parameter sets, we observe a sudden fall at neutron number N=182 for all the inputs, while in the case of 0.11, it's slightly notable. The proton pairing gap is almost the same for all the input pairing strengths. But in the case of studying the neutron pairing gap, we find a bump at neutron number N=182, reflecting the magic nature of the nucleus $^{302}120$. Again, the zero pairing gap re-confirms the magicity occurring at N=182, throughout the isotopic series of proton shell closure Z=120. For instance, the input 0.11 is showing a different behaviour in both the forces, which is probably due to over pairing estimation. 
\subsection{Symmetry energy and its co-efficients} 
\begin{figure}[H]
\centerline{\includegraphics[width=1.0 \columnwidth]{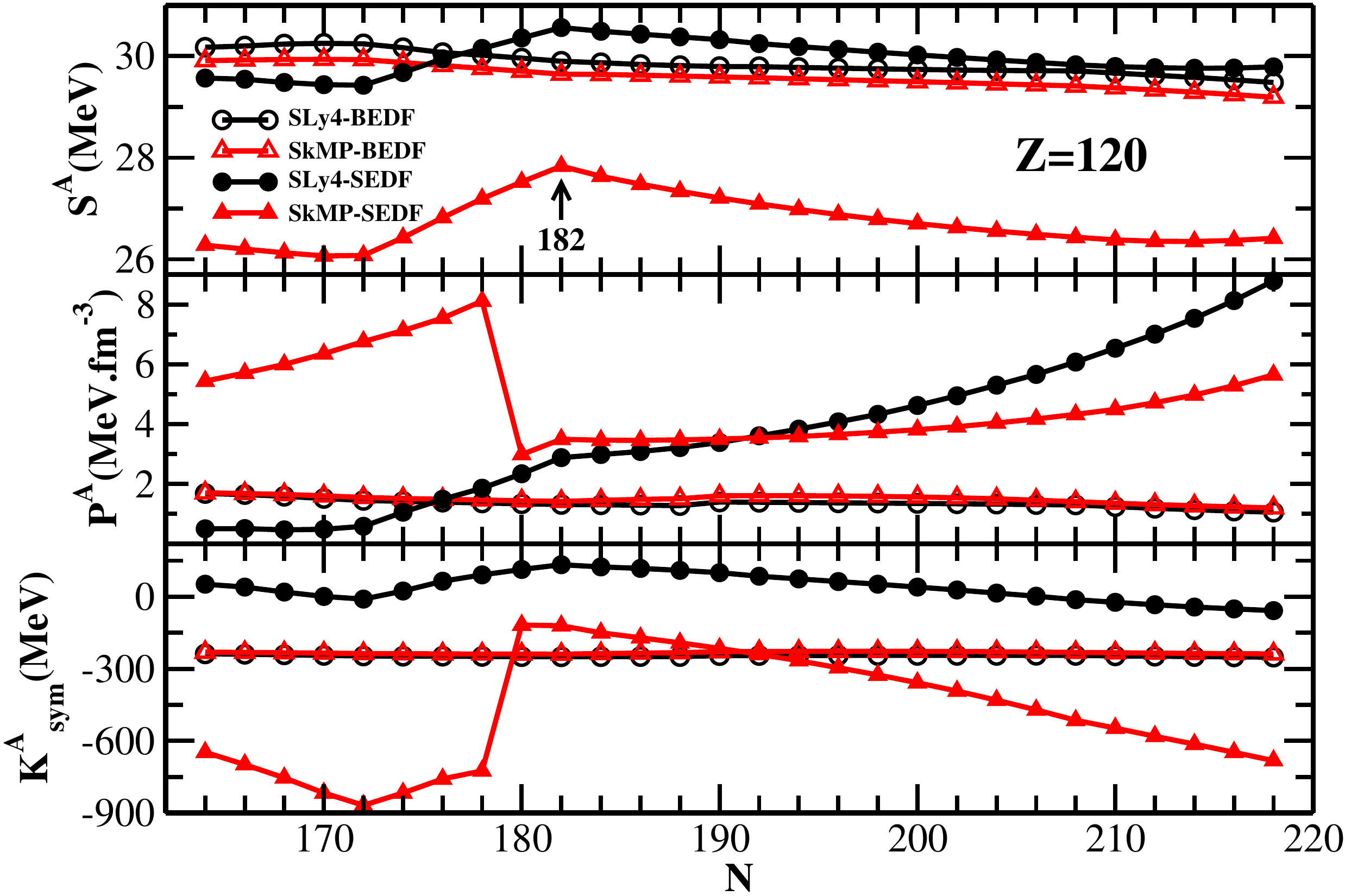}}
\vspace*{8pt}
\caption{(Color online) The nuclear symmetry energy ($S^A$), neutron pressure ($P^A$) and the curvature coefficient of symmetry energy ($K^A_{sym}$) as a function of neutron number (N) for Z=120 are presented with the non-relativistic force parameters SLy4 and SkMP.} \protect\label{fig4}
\end{figure}

In Figure \ref{fig4}, we have shown the estimated symmetry energy ($S^A$), the neutron pressure ($P^A$) and the curvature of the symmetry energy ($K_{sym}^A$) for a wide range of isotopes from neutron number N=164 to N=218 of the Z=120 nucleus within both the conventional Br\"uckner and newly derived Skyrme energy density functional for non-relativistic SLy4 and SkMP parameter sets. In a recent work, these quantities with E-RMF densities for G3 and IOPB-I sets are reported in Ref. \cite{Pat21a}. The upper panel represents the $S^A$, the middle one $P^A$ and the lower panel shows the curvature coefficient of the symmetry energy for the SHF parameters. From the upper panel of Figure \ref{fig4}, it is seen that the trend of symmetry energy ($S^A$) increases first up to N=182 and then suddenly decreases up to the drip line within the newly derived Skyrme energy density functional (SEDF) rather than Br\"uckner energy density functional (BEDF). In all three panels, the SEDF dominates over the BEDF showing an excellent agreement in results. A peak in symmetry energy is noticed at N=182 with SEDF, which is absent in the predictions of BEDF. Also there is evidence regarding the shell/sub-shell closure at N=182 or 184, which is widely accepted as the next magic combination of (neutron and proton at Z=120) and N=182/184 with SHF and E-RMF formalism \cite{Bhu12,Pat21a}.
\\
In a recent work, Pattnaik {\it et. al.} \cite{Pat22} show that the peak of the magic number in the $S^A$ is shifted to the lower isotopic number while using the Br\"uckner energy density functional (B-EDF) in the Local Density Approximation (LDA). The actual peak appears at the appropriate neutron number, once the proper fitting approach of Kumar {\it et. al.} \cite{Ank21} is applied, instead of using the Br\"uckner energy density functional. This shifting effect is correlated to the saturation properties in the nuclear matter system, i.e., the energy and momentum do not match with the empirical values in the Br\"uckner's energy density. Thus, this impact is connected to the Coester-band issue in Br\"uckner's functional.  As a result, the peak at N=172 is the peak that must appear at N=182, which just so happens to be the magic number that the E-RMF formalism predicts. Earlier the same problem is noticed for the relativistic case with G3 and NL3 sets. This problem is resolved with the works of Kumar {\it et. al.} \cite{Ank21}, where they correlated the analysis to the Coester-band problem \cite{Coe70,Coe2}, with a newly fitted relativistic energy density functional. Pattnaik {\it et. al.} \cite{Pat22} have shown a comparison between E-RMF and B-EDF approaches for Pb- isotopic series in which they have mentioned that the peak is not appearing at N=126 with the B-EDF approach.  After introducing the E-RMF fitting, they got a significant peak at N=126. Here, also a similar approach has been introduced for the non-relativistic case. A proper fitting has been done to resolve the Coester band problem in the non-relativistic case as it was done before for the relativistic case \cite{Ank21} and results are shown in Fig. \ref{fig4}, in the predictions by the newly generated SEDF. Here we can say, the results with the Skyrme energy density functional (SEDF) are fully parameter dependant. The SEDF with SLy4 set contributes a peak at N=182 in the curve of neutron pressure as well as in curvature, while the SkMP is good with the trend of neutron pressure but a little disturbed in the curve of symmetry energy curvature. The B-EDF fails to predict the results and shows almost a linearly constant trend in both the quantities. 
\begin{figure}[H]
\centerline{\includegraphics[width=1 \columnwidth]{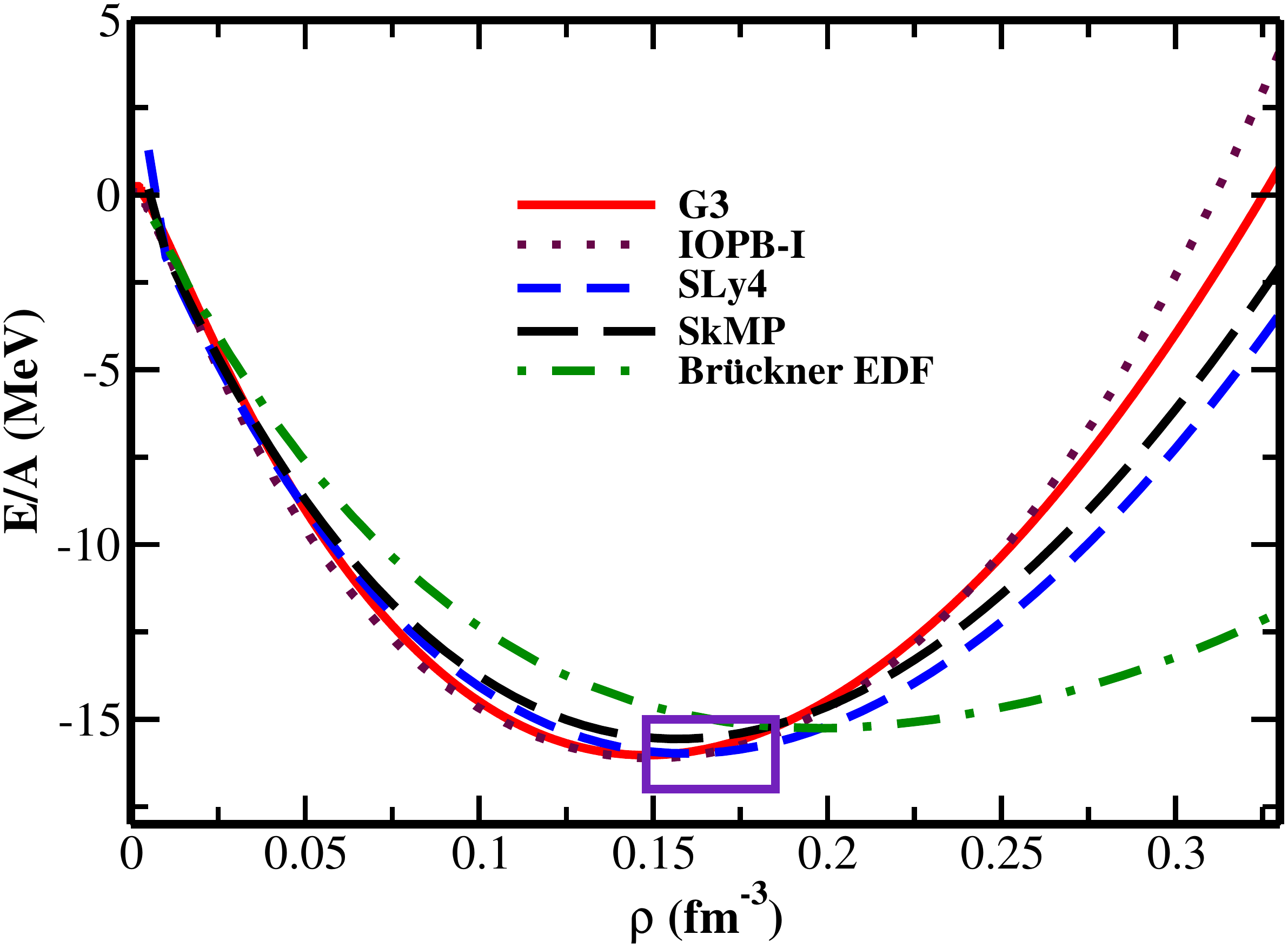}}
\vspace*{8pt}
\caption{(Color online) The nuclear matter binding energy per nucleon versus the baryon density for E-RMF (G3 solid red, IOPB-I dotted maroon) and non-relativistic Skyrme Hartree-Fock (SLy4 short-dashed blue, SkMP long-dashed black) along with the Br\"uckner energy density functional (dot-dashed green). The box-like structure defines the Coester-band.} \protect\label{fig5}
\end{figure}
\subsection{Coester-band problem}

The plot of the nuclear matter binding energy per nucleon $E/A$ as a function of the baryonic density $\rho$ for symmetric nuclear matter, which is popularly known as Coester-band. In principle, for any nucleon-nucleon potential, the calculated binding energy per particle and the baryonic density are supposed to match within the empirical limit. However, in practice, it deviates considerably from the empirical limits as shown in Figure \ref{fig5}, which is known as the Coester-band problem \cite{Coe70}. In detail, it is shown in Figure \ref{fig5} that the $E/A$ with B-EDF, the Coester-band is shifted towards the lower density region. It is noticed that the Br\"uckner's energy density functional (B-EDF) underestimates the $E/A$ value as -14.9 MeV and overestimates the baryon density ($\rho$) as 0.2 $fm^{-3}$, whereas the empirical values are $E/A$ = -16 MeV and $\rho$=0.15 $fm^{-3}$.  In the case of relativistic energy density functional (E-RMF), the deepest minimum of the curve passes through the box-like structure (Coester-band) at $E/A$=-16 MeV and baryon density $\rho$=0.15 $fm^{-3}$. Hence, the Coester-band problem issue is solved, which connects with the energy density as well as surface properties (symmetry energy, neutron pressure, curvature co-efficient) of the nucleus. After analysing this, we infer that the Coester-band problem is responsible for the disturbance/fluctuation in the appearance of peaks using the B-EDF. In Ref. \cite{Pat22}, the issue is solved successfully by taking the E-RMF functional and able to produce the peak at N=126 for Pb, which disappears/shifts with B-EDF. That is why we suggest here that a proper fitting is needed to make this functional in conjunction with the Coester-Band problem. In this regards, this present paper, first time implement the exact fitting procedure to Skyrme forces SLy4 and SkMP to derive the Skyrme energy density functional.\\
\\
\begin{figure}[H]
\centerline{\includegraphics[width=1.0 \columnwidth]{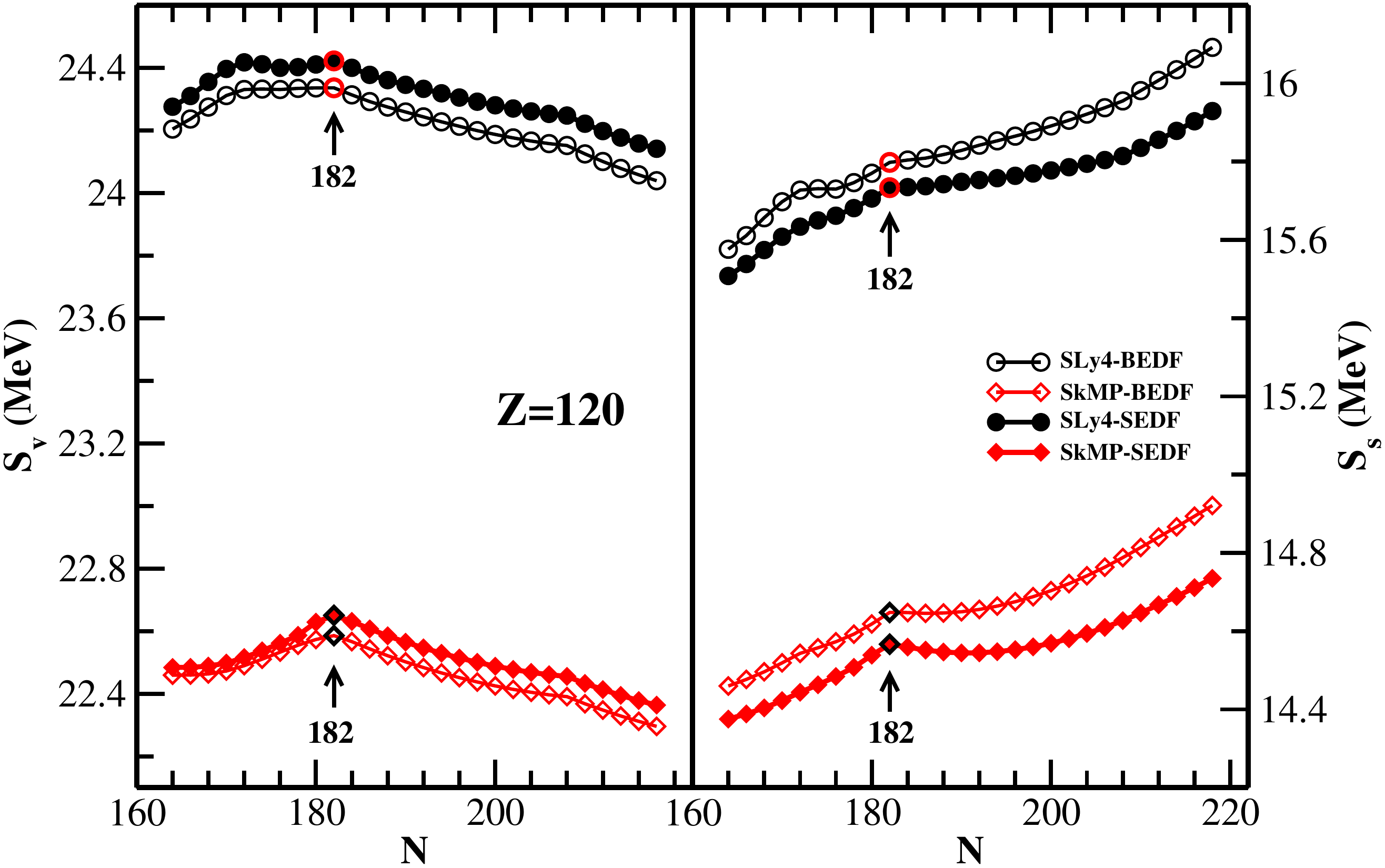}}
\vspace*{8pt}
\caption{(Color online) The volume and surface symmetry energies $S^A_v$ and $S^A_s$ are estimated for Z=120 isotopic series with the non-relativistic SLy4 and SkMP parameter sets versus neutron number N. The possible magic number N=182 is marked with an arrow.} \protect\label{fig6}
\end{figure}
\subsection{Volume and surface contribution of the symmetry energy}
The surface region of the nucleus again consists of its inner and outer surface, hence the symmetry energy is also contributed from these two portions of the nucleus. In Figure \ref{fig6}, we have estimated the volume $S^A_v$ and surface $S^A_s$ contribution of symmetry energy $S^A$ using the Danielwicz’s Liquid Drop prescription \cite{Dan03,Dan04,Dan06}. We have noticed in our recent work \cite{Pat22}, that the volume contribution of symmetry energy is higher than the surface contribution in the relativistic case with G3 and NL3 parameter sets. We find the same thing in the non-relativistic case as well but with only differences in magnitude, the volume contribution is either outpacing the surface contribution or dominating for both the functional. Also, we observe a peak at N=182 in the volume and surface symmetry energies for the SkMP parameter prominently, while it is slightly noticeable in the case of the SLy4. Similarly, for the SkMP set in the surface part, although a peak is there at N=182, it is not as prominent as in the volume part. The peak is absent for the SLy4 set. Again the study of surface and volume contribution suggests, a proper fitting is required to obtain the peak at the considered neutron number N.\\

\subsection{The total nuclear reaction cross-section}
\begin{figure}[H]
\includegraphics[width=1.0 \columnwidth]{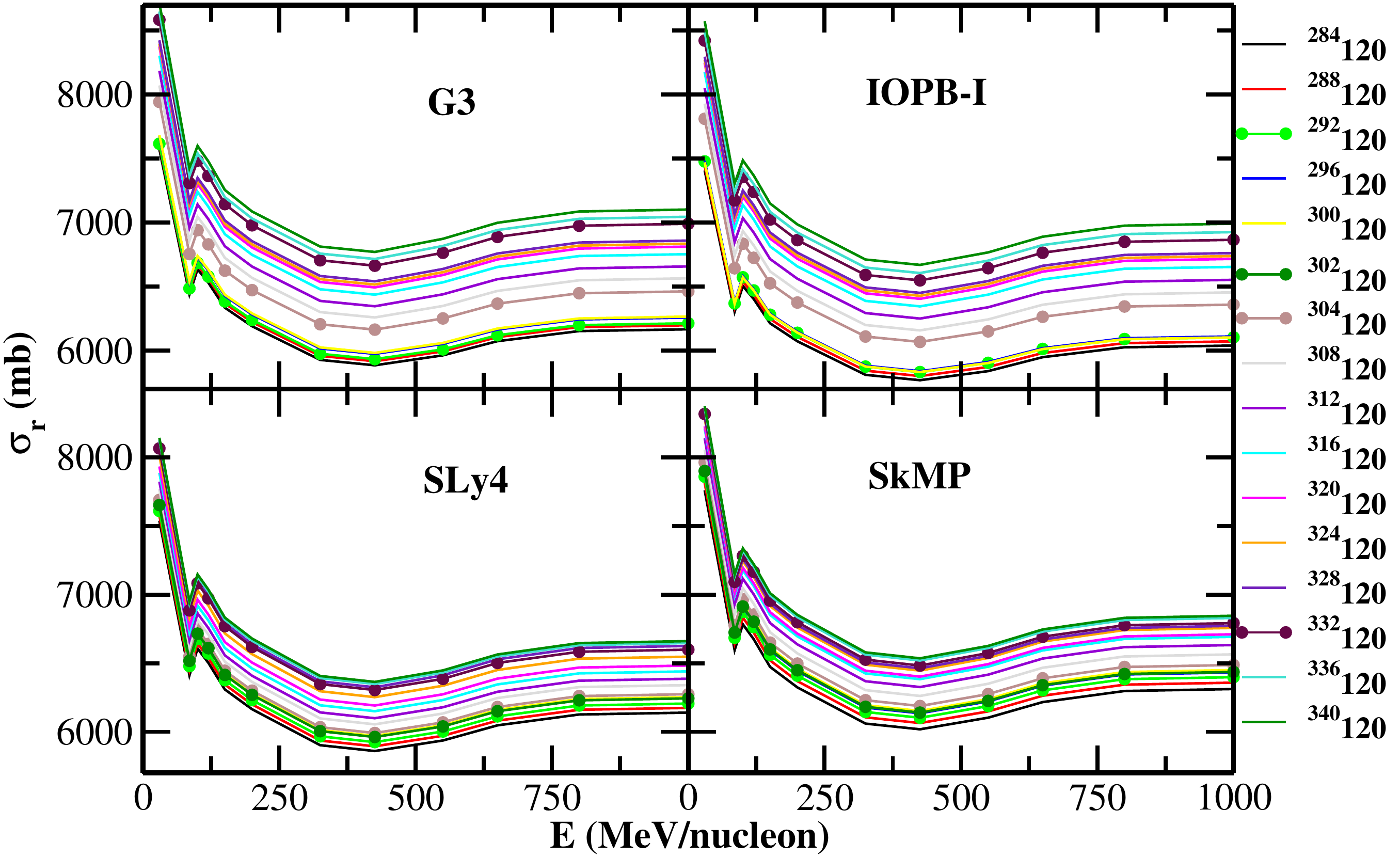}
\vspace*{8pt}
\caption{(Color online) The total nuclear reaction cross-sections $\sigma_r$ for various isotopes of Z= 120 (N = 164-220) on $^{12}$C target as a function of incident energies E.} \protect\label{fig8}
\end{figure}

In the present version of our Glauber model calculations, an analytical Gaussian form densities distribution for the target as well as the projectile is needed. Thus, the obtained axially deformed densities or the spherical densities from E-RMF or SHF are converted to their spherical equivalent densities with two Gaussian fittings. This, we have discussed in Subsection \ref{density}. Here, we get the Gaussian parameters $a_1$, $a_2$ and $c_1$, $c_2$, which are used in the Glauber model code as the inputs for the densities of the target and projectile to calculate the total nuclear reaction cross-section $\sigma_r$ and the differential elastic scattering cross-sections $\frac{d\sigma}{d\Omega}$. In this conversion process, although, we compromise to some extent the actual densities, even then we get results quite reasonable \cite{Pan11,Pan09}. The results of our calculations with E-RMF (G3 and IOPB-I) and SHF (SLy4 and SkMP) for the total nuclear reaction cross-sections are shown in Figure \ref{fig8}. The $\sigma_r$ as a function of incident energy of the projectile is plotted for these four parameter sets.\\
It has been seen in some of our earlier studies \cite{Pan11,Pan09,Pan18,Sha16} that the Glauber model works better for both spherical and deformed RMF densities and agree well with the available experimental measurements irrespective of both stable and unstable targets/projectiles. In this work, we intend to analyze the total nuclear reaction cross-section $\sigma_r$ for Z=120 isotopes. A larger $\sigma_r$ implies the formation of the combined nucleus, indicting the magic structure of the isotope. Within the Glauber model, the total nuclear reaction cross-sections ($\sigma_r$) are calculated using both non-relativistic and relativistic densities for the isotopes of Z= 120 with neutron number N=164-220 as projectiles and $^{12}$C as the target at various incident energies. The results are depicted in Figure \ref{fig8} and from which it has been found that at relatively lower incident energies around 30–200 MeV/A of the projectile, the total nuclear reaction cross section becomes maximum and then decreases rapidly with an increase in energy still around 500 MeV/A and subsequently it remains constant for further increase in energy for all the isotopes in non-relativistic and also relativistic cases respectively. Here in all of these cases the total reaction cross-sections increase with projectile mass and also the maximum value of $\sigma_r$ occurs at a particular incident energy per nucleon, irrespective of the mass of the projectiles.\\
\\
\begin{figure}[H]
\centering
\includegraphics[width=1.0 \columnwidth]{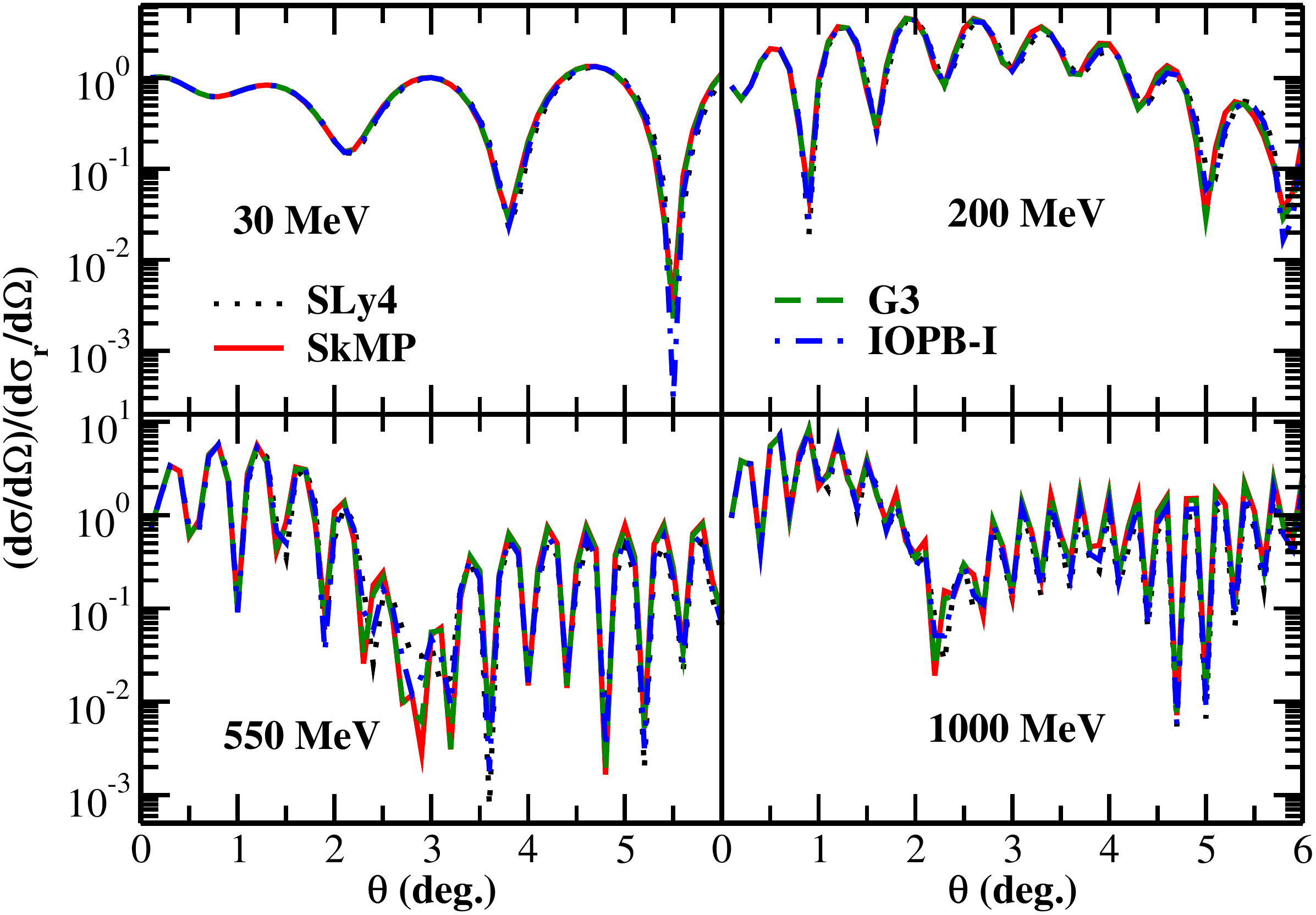}
\vspace*{8pt}
\caption{(Color online) Differential elastic scattering cross-section for $^{304}$120 + $^{12}$C systems at incident energies 30, 200, 550 and 950 MeV/nucleon as a function of scattering angle using E-RMF  (G3 dashed green, IOPB-I dot-dashed blue) and SHF (SLy4 dotted black, SkMP solid red) densities.} \protect\label{fig9}
\end{figure}
A minute investigation on total reaction cross-section in non-relativistic cases for both the parameter sets SLy4 and SkMP suggests that $^{302}$120 has higher reaction cross-sections in comparison to their neighbouring nuclei for all the incident energy. This supports the magic characteristics of the above nuclei for neutron number N = 182. The most important presumption in relativistic G3 and IOPB-I cases can be drawn from the behaviour of $\sigma_r$ for the system $^{304}$120 + $^{12}$C among all the nuclei with higher reaction cross-sections to that of their neighbouring nuclei for all the incident energy. This gives evidence of the magic number for N = 184 among all the considered isotopes of Z = 120.
\subsection{The differential elastic scattering cross-section}
The results of differential elastic scattering cross-section have a vital role in the study of scattering phenomena. Figure \ref{fig9} shows the $\frac{d\sigma}{d\Omega}$ of our calculations for the system $^{304}$120 + $^{12}$C at the incident energies 30, 200, 550 and 1000 MeV/nucleon respectively for both the relativistic and non-relativistic densities used in the Glauber model calculations, respectively.  In general, the calculated $\frac{d\sigma}{d\Omega}$ for both SLy4 and SkMP densities in non-relativistic formalism are almost similar to each other. Similarly, in the case of relativistic G3 and IOPB-I densities also, we find similar results. Thus, the differential elastic scattering cross-section in the ekinol approximation of the Glauber model is signifying the force independent nature irrespective of projectile energy and angular distribution spectrum. It is observed from the figures that the differential scattering cross-section increases significantly with the increase of scattering angles. We also noticed the dependence of the differential elastic scattering cross-section on incident energy. Furthermore, the differential cross-section shows the diffraction patterns at small values of scattering angles within $5-6^{0}$ and oscillatory behaviour at large angles. The superposition of Coulomb and nuclear amplitudes at small angles leads to unique Fresnel diffraction types of behaviour. It is also observed oscillation at higher incident energies.

\section{Summary and Conclusions}
\label{conclusion}
\noindent
In conclusion, various ground state properties such as binding energy, charge radius, neutron skin thickness, density distributions, two neutron separation energy and proton/neutron pairing gaps etc. are studied for a wide range of isotopes for Z=120 and N=164-220. All the extensive calculations are carried out by using the effective field motivated relativistic mean-field (E-RMF) as well as the non-relativistic Skyrme-Hartree-Fock formalisms. Our calculated results are compared separately for relativistic and non-relativistic cases and found that the conclusions are in general the same with small exceptions in the neutron magic number. In the E-RMF finding, N=184 is the possible neutron magic number and it is N=182 for SHF formalism. The neutron number N = 182 is a shell closure for the SHF, which is further supported by  an examination of the neutron pairing gap $\Delta_n$. The gap is virtually at the zeroth axis, which is nothing more than a magic number's property. The finite nuclear surface properties like symmetry energy, neutron pressure, and the curvature coefficient of symmetry energy as a function of neutron number within the coherent density fluctuation model with the new Skyrme energy density functional and the older Br\"uckner energy density functional are discussed. The volume and surface contributions of symmetry energy are estimated correspondly with the help of Dainelewicz's liquid drop approximation within the CDFM approach.

In the non-relativistic case, the peak in symmetry energy appeared at N=172, which is supposed to appear at N=182 in the Br\"uckner approach. This shifting of the peak in $S^A$ is correlated with the Coester-band problem in the Br\"uckner energy density functional. Similarly, the volume and surface contribution individually can produce the peak at N=182.  But combinedly the effect of both the surface and volume contribution on symmetry energy shifted the peak from N=182 to N=172. The Coester-band issue is the cause of this significant shift. By replacing the Br\"uckner's functional with an appropriate two parameter fitting formalism, Kumar {\it et. al.} \cite{Ank21} and Pattnaik {\it et. al.} \cite{Pat22} are able to resolve the peak's shifting issue.  Here also, we concluded that a proper fitting is needed to reproduce the peaks at the appropriate magic numbers for the symmetry energy. Hence, we use the same fitting procedure to establish the skyrme energy density functional for the forces SLy4 and SkMP. That is the reason why the skyrme energy density functional is totally parameter dependent. The total reaction cross-section in relativistic (G3 and IOPB-I) parameter sets suggest that $^{304}$120 have higher reaction cross-sections in comparison to their neighbouring nuclei for all the incident energy, while the same is absent with non-relativistic  (SLy4 and SkMP) at $^{302}$120. A further systematic study/better theoritical formalism in studying the reaction cross-section will be appreciated. However, this supports the magic character of the above nucleus for the neutron number N = 184 with E-RMF densities, within the Glauber model formalism despite of the discrepancy in SHF for N=182. The differential cross-section study supports the force independent nature and significant increase with the scattering angle.

\noindent 
{\bf  Acknowledgement:} JAP and KCN acknowledge Institute Of Physics (IOP), Bhubaneswar for providing the necessary computer facilities during the work. SERB (Project Nos. CRG/2019/002691 and EMR/2015/002517) partly reinforces this work. MB acknowledges the support from FOSTECT Project No. FOSTECT.2019B.04, and FAPESP Project No. 2017/05660-0.

\noindent

\end{document}